\documentclass[letter,twocolumn]{jpsj3}
\usepackage{txfonts}

\usepackage{graphicx}
\usepackage{ulem}
\usepackage{color}

\title{
Superconducting Transition Temperatures of up to 47 K from Simultaneous Rare-Earth Element and Antimony Doping of 112-Type CaFeAs$_2$
}

\author{
Kazutaka Kudo$^{1,2}$\thanks{kudo@science.okayama-u.ac.jp}, 
Yutaka Kitahama$^{1}$, 
Kazunori Fujimura$^{1}$, \\ 
Tasuku Mizukami$^{1}$, 
Hiromi Ota$^{3}$, 
and Minoru Nohara$^{1,2}$\thanks{nohara@science.okayama-u.ac.jp}
}

\inst{
$^1$Department of Physics, Okayama University, Okayama 700-8530, Japan \\
$^2$Research Center of New Functional Materials for Energy Production, Storage and Transport, Okayama University, Okayama 700-8530, Japan \\
$^3$Division of Instrumental Analysis, Department of Instrumental Analysis \& Cryogenics, Advanced Science Research Center, Okayama University, Okayama 700-8530, Japan
} 

\abst{
The effects of simultaneous Sb doping on the superconductivity of 112-type Ca$_{1-x}$$RE_x$FeAs$_2$ ($RE$ = La, Ce, Pr, and Nd) were studied through measurements of the magnetization and electrical resistivity. 
In Sb-free materials, the superconducting transition temperature $T_{\rm c}$ of the La-doped sample was 35 K, while those of the Pr- and Nd-doped samples were $\sim$10 K; no superconductivity was observed in the Ce-doped sample. 
Sb doping increased the $T_{\rm c}$ of all $RE$-doped samples: $T_{\rm c}$ increased to 47, 43, 43, and 43 K for $RE$ = La, Ce, Pr, and Nd, respectively. 
We also found that the enhanced superconductivity results from the increase in the lattice parameter $b$, which increases the As-Fe-As bond angle to be closer to the ideal tetrahedron value. 
These observations provide insight for further increasing the $T_{\rm c}$ of the 112 phase. 
}

\begin{document}
\maketitle

The discovery of iron-based superconductors has stimulated the development of novel superconducting materials\cite{Ishida,Paglione,Johnston} such as $RE$FeAsO (1111 type)\cite{Kamihara}, $AE$Fe$_2$As$_2$ (122 type)\cite{Rotter,Kudo1}, $A$FeAs (111 type)\cite{Tapp}, and FeSe (11 type)\cite{Hsu} that contain rare-earth ($RE$), alkali-earth ($AE$), and alkali ($A$) elements, as well as the development of compounds with perovskite- and/or rocksalt-type\cite{Kawaguchi,Zhu2,Ogino,Shirage} and pyrite-type spacer layers\cite{Kakiya,Nohara,Ni,Lohnert,Hieke,Kudo2}. 
For this class of materials, the maximum superconducting transition temperature $T_{\rm c}$ is 55 K,\cite{Ren}
and new iron-based superconducting materials will need to be developed to further increase $T_{\rm c}$.

To this end, novel 112-type iron arsenides of Ca$_{1-x}$La$_x$FeAs$_2$, Ca$_{1-x}$Pr$_x$FeAs$_2$, and Ca$_{1-x}$$RE_x$FeAs$_2$ ($RE$ = Ce, Nd, Sm, Eu, and Gd) reported by Katayama {\it et al.}\cite{Katayama}, Yakita {\it et al.}\cite{Yakita}, and Sala {\it et al.}\cite{Sala}, respectively, have received considerable attention, and it has been recognized that the $RE$ substitution is necessary for stabilizing the 112 phase \cite{Katayama,Yakita,Sala}. 
These compounds crystallize in a monoclinic structure with a space group of $P2_1$ (No. 4)\cite{Katayama} or $P2_1/m$ (No. 11)\cite{Yakita,Sala} and consist of alternately stacked FeAs and arsenic zigzag bond layers, which are a notable feature. 
The arsenic zigzag bond layers are considered to be composed of As$^-$ ions with a 4$p^4$ configuration as found in $RET$As$_2$ ($T =$ Ag, Au)\cite{Rutzinger}. 
Thus, the chemical formula for these compounds can be written as (Ca$^{2+}_{1-x}$$RE^{3+}_x$)(Fe$^{2+}$As$^{3-}$)As$^- \cdot xe^-$ with an excess charge of $xe^-{\rm /Fe}$ injected into the Fe$^{2+}$As$^{3-}$ layers. 
Most 112-type iron arsenides exhibit superconductivity: La-doped compounds show bulk superconductivity at 35 K\cite{Kudo3}, while Pr-, Nd-, Sm-, Eu-, and Gd-doped compounds show superconductivity at 10--15 K with a small shielding volume fraction (VF) of 5--20\%\cite{Yakita,Sala}. Ce-doped compounds rarely exhibit superconductivity\cite{Sala}.

Recently, it was reported that the simultaneous doping of isovalent P or Sb drastically improves the superconductivity in the La-doped 112 phase; $T_{\rm c}$ increased to 41 and 43 K as a result of 0.5\% P and 1\% Sb doping, respectively\cite{Kudo3}. 
In this Letter, we report that a large amount of Sb doping further increases the $T_{\rm c}$ of Ca$_{1-x}$La$_x$Fe(As$_{1-y}$Sb$_y$)$_2$ to 47 K, which is the second highest $T_{\rm c}$ after 1111-type iron-based superconductors. 
Moreover, we show that bulk superconductivity at 43, 43, and 43 K is induced by the simultaneous Sb doping of Ca$_{1-x}$$RE_x$FeAs$_2$ with $RE$ = Ce, Pr, and Nd, respectively, and that an increase in the lattice parameter $b$, which modifies the As-Fe-As bond angle, is important for optimizing the superconductivity in the 112 phase.

Single crystals of Ca$_{1-x}$$RE_{x}$Fe(As$_{1-y}$Sb$_y$)$_2$ ($RE =$ La, Ce, Pr, and Nd) were grown by heating a mixture of Ca, $RE$, FeAs, As, and Sb powders with nominal compositions of $x$ = 0.10 and $0.00 \le y \le 0.10$ (grown quantities of the 112 phase drastically decrease for $y \ge 0.20$). 
A stoichiometric amount of the mixture was then placed in an aluminum crucible and sealed in an evacuated quartz tube.
Samples were prepared in a globe box filled with argon gas. 
The ampules were heated at 700 $^\circ$C for 3 h, heated to 1100 $^\circ$C at a rate of 46 $^\circ$C/h, and then cooled to 1050 $^\circ$C at a rate of 1.25 $^\circ$C/h before furnace cooling. 
The obtained samples were characterized by powder X-ray diffraction (XRD) using a Rigaku RINT-TTR III X-ray diffractometer with Cu$K_{\alpha}$ radiation and by single-crystal XRD using a Rigaku Single Crystal X-ray Structural Analyzer (Varimax with Saturn). 
Ca$_{1-x}$$RE_{x}$Fe(As$_{1-y}$Sb$_y$)$_2$ samples were obtained together with a powder mixture of $RE$As, FeAs, FeAs$_2$, and CaFe$_2$As$_2$. The single crystals of Ca$_{1-x}$$RE_{x}$Fe(As$_{1-y}$Sb$_y$)$_2$ separated from the mixture were plate-like with typical dimensions of (0.3--0.7) $\times$ (0.3--0.7) $\times$ 0.02 mm$^3$. 
\begin{figure}[t]
\begin{center}
\includegraphics[width=7cm]{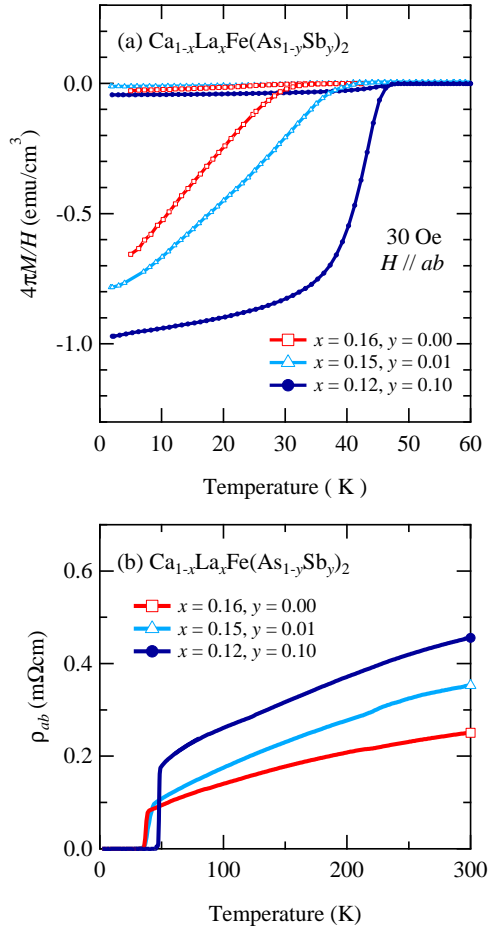}
\caption{\label{fig1}
(Color online) (a) Temperature dependence of the magnetization $M$ of Ca$_{1-x}$La$_{x}$Fe(As$_{1-y}$Sb$_y$)$_2$ measured at a magnetic field $H$ of 30 Oe parallel to the $ab$ plane under zero-field-cooling and field-cooling conditions. (b) Temperature dependence of the electrical resistivity $\rho_{ab}$ of Ca$_{1-x}$La$_{x}$Fe(As$_{1-y}$Sb$_y$)$_2$ parallel to the $ab$ plane. 
}
\end{center}
\end{figure}

The $RE$ content $x$ was analyzed by energy-dispersive X-ray spectrometry (EDS), but the Sb content $y$ could not be determined by EDS because the Sb peak positions in the EDS spectra overlapped those of Ca. 
We thus assumed a nominal $y$. 
Although the nominal $x$ was fixed at 0.1, the $x$ values measured by EDS varied with the nominal $y$. 
For example, Ca$_{1-x}$La$_{x}$Fe(As$_{1-y}$Sb$_y$)$_2$ had compositions of $x$ = 0.16, 0.15, and 0.12 with nominal $y$ = 0.00, 0.01, and 0.10, respectively; i.e., the La content tended to decrease with increasing $y$. 
In contrast, for Ca$_{1-x}$$RE_{x}$FeAs$_2$ ($RE$ = Ce, Pr, and Nd), we found compositions of $x$ = 0.15--0.25 with nominal $y$ = 0.00, 0.01, 0.05, and 0.10; there was no clear relation between $x$ and $y$.

The magnetization $M$ in a magnetic field of 30 Oe parallel to the $ab$ plane was measured using a Quantum Design magnetic property measurement system (MPMS). 
The $T_{\rm c}$ values were determined from the onset of the diamagnetism, which is characteristic of the superconducting transition. 
Electrical resistivity $\rho_{ab}$ parallel to the $ab$ plane was also measured using a standard DC four-terminal method in a Quantum Design physical property measurement system (PPMS).

%
\begin{figure}[t]
\begin{center}
\includegraphics[width=8.5cm]{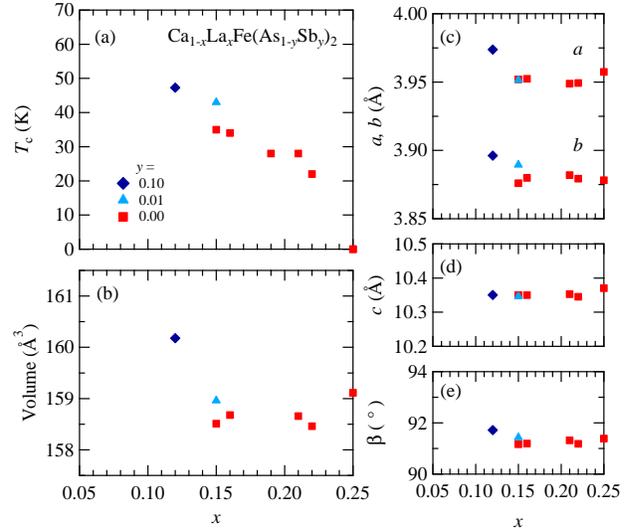}
\caption{\label{fig2}
(Color online) Dependence of the (a) $T_{\rm c}$, (b) cell volume, (c) $a$ and $b$ parameters, (d) $c$ parameter, and (e) $\beta$ angle of Ca$_{1-x}$La$_{x}$Fe(As$_{1-y}$Sb$_y$)$_2$ on $x$. 
}
\end{center}
\end{figure}
The enhancement in the superconductivity of the Sb-doped Ca$_{1-x}$La$_{x}$FeAs$_2$ can be observed in the temperature dependence of $M$ shown in Fig.~\ref{fig1}(a). 
Previous studies\cite{Katayama,Kudo3} have reported that La-doped samples with $y$ = 0.00 show diamagnetic behavior below $T_{\rm c}$ = 34 K\cite{Katayama}, and $T_{\rm c}$ increased to 43 K for $y$ = 0.01\cite{Kudo3}. 
The values of VF at the lowest temperature were previously estimated to be 66\% and 78\% for $y$ = 0.00\cite{Katayama} and 0.01\cite{Kudo3}, respectively, indicating bulk superconductivity. 
We found that a large amount of Sb doping leads to a further increase in $T_{\rm c}$ to 47 K. 
As shown in Fig.~\ref{fig1}(a), the La-doped sample with $y$ = 0.10 exhibits clear diamagnetic behavior below $T_{\rm c} =$ 47 K, and
the VF at 2 K was estimated to be approximately 100\%, which supports the emergence of bulk superconductivity. 
Further evidence of the enhanced superconductivity was obtained from the temperature dependence of $\rho_{ab}$, as shown in Fig.~\ref{fig1}(b). 
We found that $\rho_{ab}$ for the La-doped $y$ = 0.10 sample exhibited a sharp drop below 49 K, and zero resistivity was observed at 47 K; both these temperatures are much higher than those of the $y$ = 0.00\cite{Katayama} and 0.01\cite{Kudo3} samples.

The increased $T_{\rm c}$ of Ca$_{1-x}$La$_{x}$Fe(As$_{1-y}$Sb$_y$)$_2$ can be explained by two effects originating from the simultaneous Sb doping: a decrease in the La content and an increase in the cell volume. 
In Ca$_{1-x}$La$_{x}$FeAs$_2$, it is known that $T_{\rm c}$ increases with decreasing $x$ and exhibits a maximum value of 35 K at the lowest $x$ of 0.15\cite{Kudo3}, as shown in Fig.~\ref{fig2}(a); a sample with a lower $x$ could potentially have a higher $T_{\rm c}$. 
Simultaneous Sb doping allows the La content to be reduced to $x$ = 0.12, and as expected, $T_{\rm c}$ increased to 47 K. 
In general, chemical substitution modifies the number of charge carriers and induces a chemical pressure. 
The primal role of La doping is charge carrier modification because the cell volume [Fig.~\ref{fig2}(b)] and lattice parameters\cite{Kudo3} [Figs.~\ref{fig2}(c), (d), and (e)] of Ca$_{1-x}$La$_{x}$FeAs$_2$ exhibit no significant changes upon La doping owing to the similar ionic radii of Ca$^{2+}$ and La$^{3+}$. 
Thus, a decrease in the La content corresponds to a reduction in the number of charge carriers. 
Secondly, simultaneous Sb doping applies a negative chemical pressure that increases the cell volume,
as shown in Fig.~\ref{fig2}(b), because the ionic radius of Sb$^{3-}$ (Sb$^{-}$) is larger than that of As$^{3-}$ (As$^{-}$). 
The increased cell volume is a result of an increase in the in-plane lattice parameters $a$ and $b$, as shown in Fig.~\ref{fig2}(c); in contrast, the $c$ parameter hardly changes, as shown in Fig.~\ref{fig2}(d). 
In iron-based superconductors, the expansion can effectively optimize the superconductivity, which is sensitive to modifications in the crystal structure\cite{Lee,Usui,Kuroki,Mizuguchi}. 
Note that the enhancement of $T_{\rm c}$ by 0.5\% P doping in the La-doped system, which was reported in our previous article\cite{Kudo3}, should be attributed to a different mechanism because the small amount of P doping\cite{P-doping} neither reduced the La content nor changed the lattice parameters\cite{Kudo3}. However, the exact mechanism is still unclear. 
\begin{figure}[t]
\begin{center}
\includegraphics[width=8.5cm]{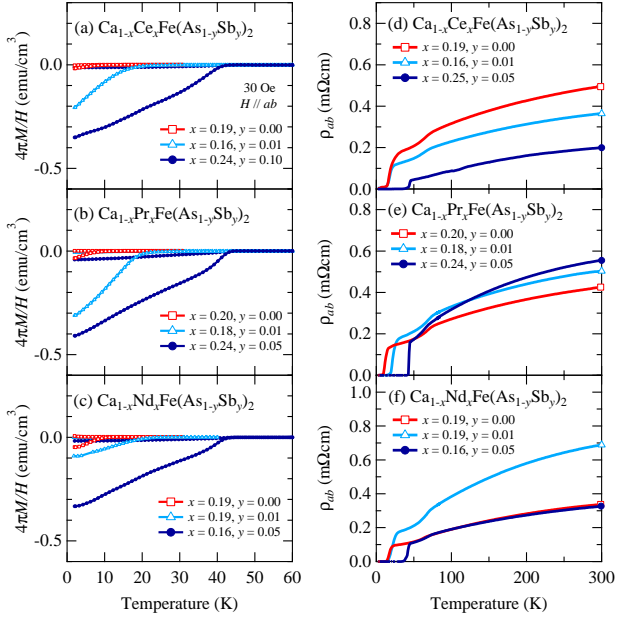}
\caption{\label{fig3}
(Color online) (a)(b)(c) Temperature dependence of the magnetization $M$ of Ca$_{1-x}$$RE_{x}$Fe(As$_{1-y}$Sb$_y$)$_2$ ($RE$ = Ce, Pr, and Nd) measured at a magnetic field $H$ of 30 Oe parallel to the $ab$ plane under zero-field-cooling and field-cooling conditions. (d)(e)(f) Temperature dependence of the electrical resistivity $\rho_{ab}$ of Ca$_{1-x}$$RE_{x}$Fe(As$_{1-y}$Sb$_y$)$_2$ ($RE$ = Ce, Pr, and Nd) parallel to the $ab$ plane. 
}
\end{center}
\end{figure}
\begin{figure}[t]
\begin{center}
\includegraphics[width=5.7cm]{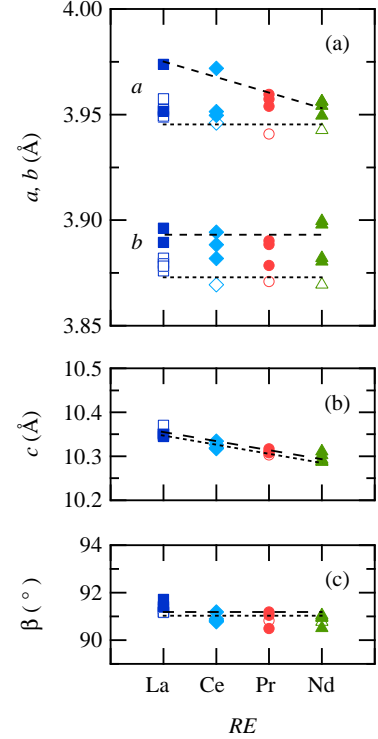}
\caption{\label{fig4}
(Color online) Dependences of the (a)  $a$ and $b$ parameters, (b) $c$ parameter, and (c) $\beta$ angle of Ca$_{1-x}$$RE_{x}$Fe(As$_{1-y}$Sb$_y$)$_2$ ($RE$ = La, Ce, Pr, and Nd) on $RE$. 
The open and closed symbols indicate the lattice parameters of Sb-free samples and Sb-doped samples, respectively. 
The dotted and broken lines are, respectively, guides to the eye for the lattice parameters of Sb-free samples and Sb-doped samples with the highest $T_{\rm c}$ for each $RE$-doped systems. 
}
\end{center}
\end{figure}

A similar enhancement in the superconductivity was also observed in Sb-doped Ca$_{1-x}$$RE_{x}$FeAs$_2$ ($RE$ = Ce, Pr, and Nd).
As shown in Figs.~\ref{fig3}(a), (b), and (c), in the Sb-free samples, the Ce-doped system shows no bulk superconductivity down to 2 K, while the Pr- and Nd-doped systems exhibit superconductivity at 10 and 11 K, respectively, with a small VF of 5\%. 
These results are consistent with previous reports\cite{Yakita,Sala}. 
Sb doping resulted in higher $T_{\rm c}$ values of 21 and 43 K in Ce-doped systems of $y$ = 0.01 and 0.10, respectively, 26 and 43 K in Pr-doped systems of $y$ = 0.01 and 0.05, respectively, and 24 and 43 K in Nd-doped systems of $y$ = 0.01 and 0.05, respectively. 
Thus, we found that Sb-doped 112 phase samples have a $T_{\rm c}$ of higher than 40 K irrespective of $RE$. 
More importantly, Sb substitution resulted in a substantial increase in the VF, indicating the emergence of bulk superconductivity. 
Evidence of the enhanced superconductivity was also found in the temperature dependence of $\rho_{ab}$. 
As shown in Figs.~\ref{fig3}(d), (e), and (f), $\rho_{ab}$ of Ca$_{1-x}$$RE_{x}$Fe(As$_{1-y}$Sb$_y$)$_2$ ($RE$ = Ce, Pr, and Nd) with $y$ = 0.05 was zero at 37, 43, and 37 K, respectively; these temperatures are much higher than those (5, 9, and 13 K) of the Sb-free samples. 
Note that the zero resistivity observed in the Sb-free Ca$_{1-x}$Ce$_{x}$FeAs$_2$ sample is attributed to filamentary superconductivity because there is no visible diamagnetic signal at $T_{\rm c}$. 
In Ca$_{1-x}$$RE_{x}$Fe(As$_{1-y}$Sb$_y$)$_2$ ($RE$ = Ce, Pr, and Nd), while we have observed the importance of the volume effect [Figs.~\ref{fig4}(a) and (b)] in the enhanced superconductivity in the same manner as discussed for the La-doped system, the precise $x$ and $y$ dependences of the $T_{\rm c}$ are still unclear.

On considering the well-known relation between $T_{\rm c}$ and local structure in the iron-based superconductors\cite{Lee,Usui,Kuroki,Mizuguchi}, the most important factor that determines the volume effect, which enhances superconductivity, is the increase in the $b$ parameter. 
In general, iron-based superconductors with high $T_{\rm c}$, such as 1111-type superconductors with $T_{\rm c} > 50$ K, satisfy the following structural conditions: the Fe and As atoms form an ideal tetrahedral structure (i.e., the As-Fe-As bond angle $\alpha$ = 109.47$^\circ$)\cite{Lee,Usui} and the As height with respect to the Fe plane, $h_{Pn}$, is approximately 1.38 \AA\cite{Kuroki,Mizuguchi}. 
Because of the monoclinic structure, the 112 phase possesses two bond angles: $\alpha_{a}$ and $\alpha_{b}$, which correspond to the $\alpha$-angle along the $a$ and $b$ axes, respectively. 
Our preliminary structural analysis\cite{Structure} of Ca$_{1-x}$La$_{x}$Fe(As$_{1-y}$Sb$_y$)$_2$ with $T_{\rm c}$ = 47 K showed that $\alpha_{b}$ is drastically improved by Sb doping: $\alpha_{a}$ = 109.098$^\circ$ and $\alpha_{b}$ = 107.061$^\circ$ in the Sb-free sample\cite{Katayama}, while $\alpha_{a}$ = 109.731$^\circ$ and $\alpha_{b}$ = 108.655$^\circ$ in the Sb-doped sample. 
The result suggests that the increase in $b$ parameter [Fig.~\ref{fig4}(a)], which increases $\alpha_{b}$, is a key factor for enhancing the superconductivity in the 112 phase\cite{a_parameter}. 
On the other hand, Sb doping does not modify $h_{Pn}$ as much, because the $c$ parameter is almost insensitive to Sb doping [Fig.~\ref{fig4}(b)]. 
In both Sb-free and Sb-doped samples, the $h_{Pn}$ values are slightly larger than the ideal value: $h_{Pn} =$ 1.418 and 1.408 \AA \ in the Sb-free\cite{Katayama} and Sb-doped samples, respectively. 
We expect that the $T_{\rm c}$ of the 112 phase can be increased above 50 K, if the $b$ parameter can be increased to approximately equal $a$ and, simultaneously, the $c$ parameter can be slightly decreased.

In summary, the effects of Sb doping on the $T_{\rm c}$ of 112-type Ca$_{1-x}$$RE_x$FeAs$_2$ ($RE$ = La, Ce, Pr, and Nd) were studied through measurements of the magnetization and electrical resistivity. 
Sb-doping results in an increase in $T_{\rm c}$ such that the $T_{\rm c}$ values of Sb-doped Ca$_{1-x}$$RE_x$FeAs$_2$ with $RE$ = La, Ce, Pr, and Nd were 47, 43, 43, and 43 K, respectively. 
We found that an increase in the $b$ parameter, which improves the As-Fe-As bond angle, is important for enhancing the superconductivity in the 112 phase.

\begin{acknowledgments}
This work was partially supported by Grants-in-Aid for Scientific Research (B) (26287082) and (C) (25400372) from the Japan Society for the Promotion of Science (JSPS) and the Funding Program for World-Leading Innovative R\&D on Science and Technology (FIRST Program) from JSPS. 
\end{acknowledgments}


\begin{thebibliography}{99}
\bibitem{P-doping} By doping P in a range of more than 1\%, the La doped 112 phase was not obtained.
\end{thebibliography}
\end{document}